\documentclass{amia}
\usepackage{amsmath, amssymb, amsfonts}
\usepackage{graphicx}
\usepackage[colorlinks=true,linkcolor=blue,citecolor=black,urlcolor=blue]{hyperref}
\usepackage{tikz}
\usetikzlibrary{shapes.geometric, arrows.meta, positioning, calc, fit, backgrounds}
\usepackage{booktabs}
\usepackage[superscript,nomove]{cite}
\usepackage[labelfont=bf]{caption}
\usepackage{tabularx}
\usepackage{booktabs}
\usepackage{array}
\usepackage{ragged2e}
\usepackage{wrapfig}
\usepackage{url}

\begin{document}

\title{Scaling Up Formal Representation of Clinical Trial Protocols in Ensemble Logic Using LLMs: A Preliminary Study}

\author{Yan Huang, PhD\textsuperscript{1}\footnotemark[1], Xubing Hao, PhD\textsuperscript{1}\footnotemark[1], Xiaojin Li, PhD\textsuperscript{1}, 
Rashmie Abeysinghe, PhD\textsuperscript{1}, Xiaoqian Jiang, PhD\textsuperscript{2}, Licong Cui, PhD\textsuperscript{1,2}, Guo-Qiang Zhang, PhD\textsuperscript{1,2}\footnotemark[2]}

\institutes{
 \textsuperscript{1}McGovern Medical School,
 \textsuperscript{2}McWilliams School of Biomedical Informatics,
 The University of Texas Health Science Center at Houston, Houston, Texas, USA
 }

\maketitle
\footnotetext[1]{These authors contributed equally to the first authorship .}
\footnotetext[2]{Corresponding author.}

\noindent{\bf Abstract}

\textit{The reliance on unstructured free text for documenting clinical trial protocols creates a significant barrier to automated reasoning, cohort discovery, and trial simulation. The lack of formal structure obscures critical temporal phenotypes, such as dynamic eligibility criteria and event timing constraints. Although Temporal Ensemble Logic (TEL) offers an expressive framework for modeling these elements, manual encoding remains a prohibitive bottleneck. We introduce the CT-TEL workflow: a scalable pipeline leveraging Large Language Models (LLMs) to translate narrative clinical protocols into TEL formulas. We applied CT-TEL to generate logical models for 23 real-world trials from ClinicalTrials.gov. We evaluated translation fidelity via a back-translation approach, using LLMs to convert TEL formulas back into natural language and measuring semantic similarity against source texts. The resulting semantic retention suggests that LLMs may offer a pathway for mapping informal protocols to computable logic, providing preliminary evidence toward scalable clinical trial emulation within the emerging ``Symbolic Biomedicine'' paradigm.}

\section{Introduction}
Real-world data (RWD), including electronic health records (EHRs), electrophysiological time-series, and other continuously recorded sources, offer unprecedented opportunities for generating real-world evidence. Formal approaches in the life sciences follow two primary modeling strategies: a mechanistic strategy that models underlying biological causes, and an observational strategy that uncovers patterns and regularities in event sequences across cohorts \cite{qel}. This work adopts the observational perspective, employing logic as a description language for specifying population subgroups and encoding temporal reasoning about clinical trials \cite{time_tel_cts}.

\begin{wrapfigure}{R}{0.30\textwidth}
\vspace{-15pt}
\centering
\begin{tikzpicture}[
    auto,
    block/.style={rectangle, draw=blue!50, rounded corners=4pt, fill=blue!5, text width=2.2cm, text centered, minimum height=1.0cm, font=\small\bfseries},
    telblock/.style={diamond, aspect=2.2, draw=green!60!black, fill=green!5, text width=1.5cm, text centered, font=\small\bfseries\itshape, inner sep=2pt},
    line/.style={draw, -{Stealth[scale=1.0]}, thick, black!70},
    eval/.style={draw, -{Stealth[scale=1.0]}, thick, dashed, red!70}
]
    \node [block] (source) {Source:\\Clinical Trials};
    \node [telblock, below=0.6cm of source] (tel) {Ensemble Logic};
    \node [block, below=0.6cm of tel] (target) {Target:\\Reconstructed\\Trials};

    \path [line] (source) -- node [right, font=\footnotesize] {Step 1: LLM} (tel);
    \path [line] (tel) -- node [right, font=\footnotesize] {Step 2: LLM} (target);
    \path [eval] (target.west) -- ++(-0.7,0) |- node [pos=0.25, left, font=\footnotesize\color{black}, text width=1.4cm, align=right] {Step 3:\\Evaluate} (source.west);
\end{tikzpicture}
\caption{LLM-driven paradigm: forward translation to TEL, reverse translation, and evaluation.}
\label{fig:paradigm}
\vspace{-8pt}
\end{wrapfigure}
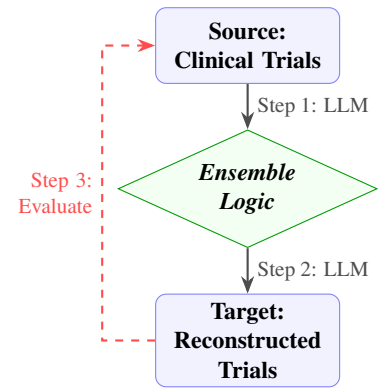

A central challenge for such observational modeling stems from clinical trial protocols. These protocols define eligibility criteria, treatment regimens, follow-up schedules, and outcome assessments. 
Each module is inherently temporal, unfolding across specific intervals and involving complex logical dependencies.
Because these constraints are expressed in unstructured free text, they resist direct computational interpretation: temporal phenotypes such as dose-escalation windows, washout periods, and event-anchored follow-up schedules cannot be directly queried, simulated, or reused across studies~\cite{weng2010formal}.
TEL \cite{zhang2024temporal} and Rational Ensemble Logic (QEL) \cite{qel} were introduced to fill this representation gap. Building upon TEL with ensemble logic, QEL defines Temporal Event Structures (TES) in the rational number domain, aiming at applications for cohort selection queries. These formalisms integrate first-order constructs with metric modal operators tailored for capturing observational temporal phenotypes. While expressive, they face a practical bottleneck: manually translating thousands of complex clinical trials into TEL is labor-intensive and error-prone.

This paper introduces a highly scalable approach: utilizing LLMs to bridge the gap between narrative clinical trial protocols and formal TEL representations. By automating the forward translation of ClinicalTrials.gov texts into TEL and rigorously validating the semantic fidelity via LLM reverse-translation, we present a robust workflow that maps real-world protocols to computable logical structures. 
Figure~\ref{fig:paradigm} illustrates our core paradigm. First, LLMs are utilized to translate unstructured source clinical trials into formal TEL. Next, to ensure our formalization is valid and strictly retains the original meaning, LLMs are used again to reverse-translate these logic formulas back into target reconstructed clinical trials. Finally, an evaluation step assesses the semantic consistency between the reconstructed texts and original sources. This cyclic paradigm makes the critical role of LLMs explicit: driving both the scalable formalization (forward translation) and the rigorous empirical validation (reverse translation).

\section{Background}

\textbf{Clinical Trials and Temporal Dynamics.}
Clinical trials involve intricate temporal dependencies among interventions, patient responses, and observational outcomes. A complete Clinical Trial Simulation (CTS) framework encapsulates both timeline-based structures and cohort-based structures \cite{time_tel_cts}. The timeline revolves around anchor points such as the Start Event ($S$) and Index Event ($I$). Between these events lies an Observation Window ($OW$), which evaluates pre-existing conditions. Eligible participants then undergo a Baseline Visit ($B$), followed by an Intervention Phase ($T$), scheduled Follow-Up Visits ($F$), and a temporally bounded Outcome Assessment ($O$).

\textbf{Logical Representation of Clinical Trials.}
The push for standardized clinical research has yielded frameworks such as the CDISC Study Data Tabulation Model (SDTM) and the BRIDG domain information model \cite{cdisc}. Furthermore, semantic web technologies like OWL and the OBO Foundry enable the standardization of concepts \cite{obofoundry, weng2010formal}. 
Prior work has applied formal temporal logic to the representation of clinical rules. Narrative eligibility criteria have been formalized using Linear Temporal Logic (LTL \cite{pnueli1977temporal}) mapped to the HL7 RIM \cite{madkour2016ltl, peleg2001hl7}, and normative concepts such as obligations and permissions have been modeled using deontic logic \cite{otte2022bfo}. However, LTL and Allen’s Interval Algebra \cite{allen1983maintaining} lack the expressiveness needed to capture the quantitative metric bounds and dense temporal patterns present in real-world data, whereas more expressive interval logics, such as Halpern-Shoham logic \cite{halpern1991propositional}, are computationally undecidable for direct model-checking applications.

\textbf{Ensemble Logic (TEL and QEL).}
Ensemble Logic \cite{qel, zhang2024temporal} is a temporal logic designed to express the kinds of time-bound clinical statements that appear in trial protocols, EHR-based phenotype definitions, and electrophysiological recordings. Two variants are used: TEL over $\mathbb{Z}^+$, whose timeline is counted in discrete units (days, weeks, months), and QEL, which extends TEL when the timeline is defined over the rational number domain. For clarity and ease of reading, all examples presented in this paper use TEL over $\mathbb{Z}^+$, whereas QEL is generally preferred for computation.

The core idea is that every clinical concept (e.g., ``diagnosis of Alzheimer's Disease'') is represented as an \textit{atomic proposition}~$p$. A formula then combines propositions with five mechanisms that mirror how clinicians describe temporal requirements in natural language:

\begin{itemize}
    \setlength\itemsep{1pt}
    \item \textbf{Shift} ($p_{x+t}$): Places a proposition at a specific offset from a reference time-point. Example: ``MMSE is assessed at week~24 after randomization'' is formalized as $(\mathit{MMSE})_{r+24w}$.
    \item \textbf{Box} ($\Box_t\, p$): Asserts that a condition holds continuously for duration~$t$. Example: ``Donepezil administered daily for 52 weeks'' as $\Box_{52w}\,\mathit{DONEPEZIL}$.
    \item \textbf{Diamond} ($\Diamond_t\, p$): Asserts that an event occurs \textit{at least once} inside a bounded window~$t$. Example: ``Follow-up visit within 4 weeks'' as $\Diamond_{4w}\,\mathit{FOLLOW\_UP}$.
    \item \textbf{Boolean connectives} ($\land$, $\lor$, $\neg$): Combine or negate propositions (e.g., exclusionary criteria). Example: ``Age $\geq$ 50 \textbf{and} no history of stroke in the past 26 weeks'' as $(\mathit{AGE\_GE50})_x \;\land\; (\Diamond_{26w}\,\neg\mathit{STROKE})_x$.
    \item \textbf{Quantification} ($\exists x$, $\forall x$): Introduces time variables so different parts of a formula refer to the same anchor. Example: ``There is an eligible screening date'' as $\exists x\;[(\mathit{SCREENING})_x]$.
\end{itemize}

Because every formula is built from these five mechanisms, the mapping between free-text protocol statements and logic is \textit{bidirectional}: a protocol sentence can be compiled into a TEL formula, and a TEL formula can be read back as an unambiguous clinical narrative. This property is exploited by the CT-TEL workflow (Section~3) for both automated formalization and rigorous reverse-translation evaluation.

\section{Methods}
\begin{figure}[h!]
\centering
\begin{tikzpicture}[
    node distance=0.75cm and 0.9cm,
    box/.style={rectangle, draw, line width=0.7pt, text width=3.4cm, align=center,
                rounded corners=5pt, minimum height=1.0cm, font=\small},
    A/.style={box, fill=cyan!10,   draw=cyan!55},
    B/.style={box, fill=blue!9,    draw=blue!50},
    C/.style={box, fill=green!14,  draw=green!55, line width=1pt},
    D/.style={box, fill=violet!9,  draw=violet!55},
    E/.style={box, fill=orange!16, draw=orange!65, line width=1pt},
    arr/.style={-{Stealth[scale=1.05]}, line width=0.85pt, black!65},
    darr/.style={-{Stealth[scale=1.05]}, line width=0.85pt, dashed, red!75},
    lbl/.style={font=\scriptsize\itshape, black!55},
]

\node (s1) [A] {
    \textbf{1.~Protocol}\\\textbf{Download}\\[2pt]
    \scriptsize 6 protocol modules
    \\\scriptsize clinicaltrials.gov};
\node (s2) [B, right=0.9cm of s1] {
    \textbf{2.~Semantic}\\\textbf{Mapping}\\[2pt]
    \scriptsize Ontology mapping\\\scriptsize $\to$ event extraction};
\node (s3) [C, right=0.9cm of s2] {
    \textbf{3.~TEL}\\\textbf{Formalization}\\[2pt]
    \scriptsize Propositions \scriptsize + logic blocks \\\scriptsize $\to$ canonical formula};

\node (lib) [
    rectangle, rounded corners=6pt, line width=1.1pt,
    draw=orange!65, fill=yellow!13,
    minimum width=13.0cm, minimum height=1.0cm,
    align=center, font=\small,
    below=0.75cm of s2
] {
    \textbf{CT-TEL Library}\\[4pt]
    \scriptsize
    Extracted Terms
    $\vert$ Defined Events
    $\vert$ Propositions 
    $\vert$ Logic Blocks
    $\vert$ Canonical Formulas
};

\node (s5) [E, below=0.75cm of lib, xshift=-2.3cm] {
    \textbf{5.~Fidelity}\\\textbf{Evaluation}\\[2pt]
    \scriptsize Compare 6 modules\\\scriptsize vs.~original text};
\node (s4) [D, right=0.9cm of s5] {
    \textbf{4.~TEL-to-}\\\textbf{Protocol}\\[2pt]
    \scriptsize Reconstructs\\\scriptsize 6 protocol modules};

\begin{pgfonlayer}{background}
    \node [rectangle, rounded corners=9pt, inner sep=8pt,
           draw=cyan!50, fill=cyan!4, line width=0.9pt,
           fit=(s1)(s2)(s3),
           label={[font=\small\bfseries, text=cyan!70]above:{Library Builder Phase}}]
          {};
    \node [rectangle, rounded corners=9pt, inner sep=8pt,
           draw=orange!55, fill=orange!4, line width=0.9pt,
           fit=(s4)(s5),
           label={[font=\small\bfseries, text=orange!75]above:{Evaluation Phase}}]
          {};
\end{pgfonlayer}

\draw[arr] (s1) -- node[above, lbl]{JSON}        (s2);
\draw[arr] (s2) -- node[above, lbl]{event array} (s3);

\draw[arr] (s2.south) --
    node[left, lbl, pos=0.7]{terms \& events}
    (lib.north -| s2.south);
\draw[arr] (s3.south) --
    node[right, lbl, pos=0.7]{props \& formula}
    (lib.north -| s3.south);

\draw[arr] (lib.south -| s4.north) --
    node[right, lbl, pos=0.3]{formula \& terms}
    (s4.north);

\draw[arr] (s4.west) --
    node[above, lbl]{\shortstack{reconst.\\6 modules}}
    (s5.east);

\draw[darr] (s1.west) -- ++(-0.8, 0)
    |- node[pos=0.32, left, font=\scriptsize\itshape, text=red!80]{source text}
    (s5.west);

\end{tikzpicture}
\caption{The CT-TEL Workflow comprises three structural components.}
\label{fig:workflow}
\end{figure}
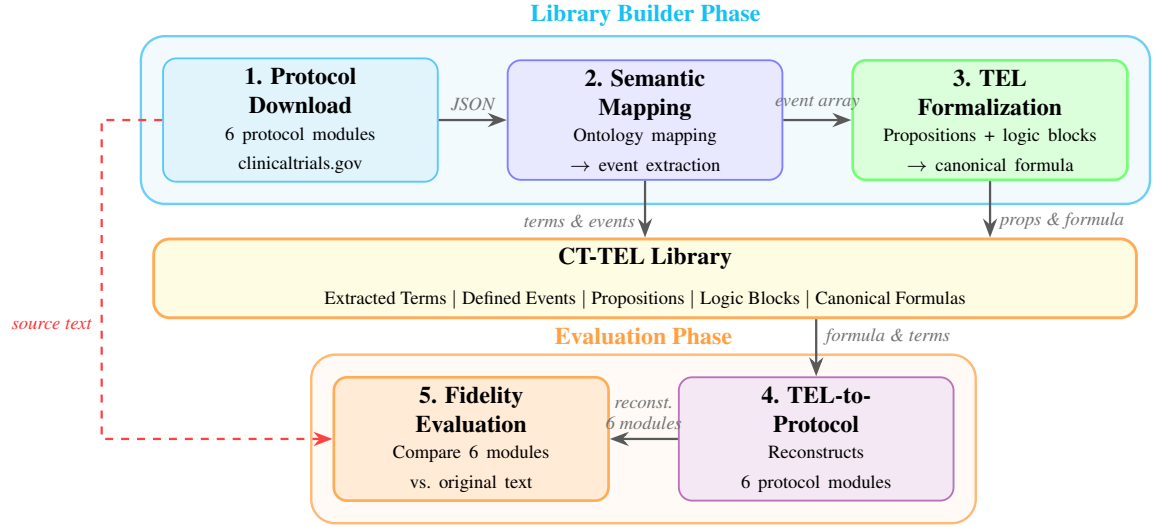

To scale the generation of formal specifications, we designed a structured experimental workflow denoted as CT-TEL, organized into three functional components: the Library Builder Phase, the central CT-TEL Library, and the Evaluation Phase, as illustrated in Figure~\ref{fig:workflow}. 
The Library Builder Phase (Steps 1-3) ingests raw JSON records from ClinicalTrials.gov, enriches clinical entities via multi-ontology semantic mapping, and formalizes the resulting event array into canonical TEL components through a four-step deterministic process (propositionalization, temporal anchor establishment, modal operator assignment, and modular synthesis). 
The Evaluation Phase (Steps 4-5) reconstructs a ClinicalTrials.gov-style protocol from the stored canonical formula and assesses fidelity via a hybrid framework of human expert review, LLM-based assessment, and algorithmic lexical and semantic scoring. All intermediate and final artifacts are committed to the CT-TEL Library at each stage, preserving full backward traceability from canonical formula to source text.

\textbf{Data Sources and Protocol Download.}
The CT-TEL workflow begins by retrieving trial protocol records directly from ClinicalTrials.gov. Each record is ingested in its native JSON format and includes six modules relevant to clinical events and temporal logic: descriptionModule, conditionsModule, designModule, armsInterventionsModule, outcomesModule, and eligibilityModule. These raw records serve as the ground-truth source for all subsequent formalization and evaluation procedures.

\textbf{Semantic Mapping and Temporal Event Extraction.}
The second phase comprises two sequential sub-tasks executed by the LLM via structured prompt engineering. 
To ensure methodological reproducibility, the exact LLM system configurations and user prompts used across all pipeline stages including entity mapping, event extraction, formula synthesis, and back-translation are publicly available in the project repository~\cite{cttel2026github}.
 Furthermore, the framework enforces structured output constraints (JSON schema definitions) during every LLM inference call. This guarantees that extracted events, structural roles, and temporal conditions are generated in a predictable format that automated downstream scripts parse sequentially without relying on fragile string matching.

\textbf{Ontological semantic mapping.}
The LLM processes each JSON record using in-context learning to identify named clinical entities: diseases, drugs, procedures, demographics, and biomarkers. The LLM is prompted to link extracted strings to Unified Medical Language System (UMLS)~\cite{umls} concept identifiers. Because zero-shot entity linking tends to hallucinate identifiers and mishandle ambiguous acronyms, the output is augmented with the Alzheimer's Disease Common Data Element Ontology for Clinical Trials (AD-CDO)~\cite{adcdo}, a domain-specific reference terminology for Alzheimer's disease clinical trials, which constrains the mapping vocabulary.
Each mapping receives a heuristic confidence label: HIGH, MEDIUM, or LOW. Domain experts review the generated JSON artifacts and manually correct any mappings flagged LOW or MEDIUM before the data progresses to the next stage. This verification step ensures that variables in downstream logical formulas are grounded in standardized medical terminologies, keeping the CT-TEL Library consistent and traceable across all 23 trials.

\textbf{Temporal event extraction.}
Building on the enriched record, the LLM parses each protocol section to isolate clinical events and their associated temporal constraints (e.g., ``within 6 months,'' ``prior to screening,'' ``at week 13''). Each extracted event is assigned to its TEL structural role defined by the CTS framework \cite{time_tel_cts}: Start Event ($S$), Index Event ($I$), Observation Window ($OW$), Inclusion Criterion ($IC$), Exclusion Criterion ($EC$), Baseline ($B$), Intervention ($T$), Follow-Up ($F$), or Outcome ($O$). The output is a structured JSON array in which each record specifies five fields: \texttt{event\_name} (a concise standardized label), \texttt{original\_text} (the verbatim source snippet for provenance), \texttt{temporal\_constraint} (the explicit temporal condition), \texttt{structural\_role}, and \texttt{ontology} (the mapped code from the mapping step). The validated JSON array output from this stage functions as the direct programmatic input for the formal logic engineer prompt in the subsequent phase, transferring the extracted variables without manual re-entry. This structured intermediate representation bridges unstructured protocol narratives and the precise syntactic requirements of temporal logic.
The extracted terms and defined events from both sub-tasks are committed to the CT-TEL Library, forming the ontology-grounded event vocabulary used by all downstream steps.

\textbf{Formalization into Temporal Ensemble Logic.}
The extracted event arrays are converted into canonical TEL formulas through a four-step process, executed by the LLM prompted as a formal logic engineer.
All propositions, logic blocks, and canonical formulas are deposited in the CT-TEL Library.

\textbf{Step 1: Propositionalization.} Each distinct clinical event, condition, or intervention is assigned a concise mnemonic label (e.g., \texttt{SURGERY}, \texttt{DRUG\_TX}), producing a formal data dictionary of atomic propositions that decouples logical reasoning from surface-level linguistic variation.

\textbf{Step 2: Temporal Anchors.} The pipeline introduces existentially quantified variables as a relative coordinate system: \(x\) marks the Start Event (e.g., screening) and \(r = x + \delta\) marks the Index Event (e.g., randomization), where $\delta$ is the protocol-defined offset ($\delta = 0$ when no explicit delay is specified).

\textbf{Step 3: Modal Operator Assignment.} Each temporal constraint is mapped to the appropriate metricized operator (cf.\ Section~2.3): $\Box_t$ for sustained states (e.g., a treatment phase of duration $t$), $\Diamond_t$ for events required within a bounded window, and $\neg\Diamond_t$ for exclusionary lookback periods.

\textbf{Step 4: Modular Synthesis.} The propositions and modal sub-formulas are assembled into conjunctive blocks. In most cases, the final formula contains four main components: Eligibility $\varphi_{\text{elig}}$ (prior to/at $x$), Baseline $\varphi_{\text{base}}$ (at $x$ or $r$), Intervention $\varphi_{\text{tx}}$ (from $r$), and Outcome $\varphi_{\text{out}}$ (at protocol offsets from $r$) where $x$ usually corresponds to the screening visit and $r$ to index event such as randomization.

\textbf{TEL-to-Protocol Translator.}
In an isolated, context-cleared session, an LLM configured as a clinical trial methodologist receives only the CT-TEL Library content (propositions, anchors, and canonical formula) and reconstructs a ClinicalTrials.gov-style protocol across six modules (descriptionModule, conditionsModule, designModule, armsInterventionsModule, outcomesModule, eligibilityModule), without access to the source text. The prompt enforces direct mapping of each TEL operator to natural language: temporal anchors to chronological phrasing, $\Box_t$ to sustained-state descriptions, $\Diamond_t$ to bounded-window descriptions, and negation to exclusionary clinical phrasing, with all proposition labels expanded into full clinical descriptions.

\textbf{Evaluation of Translation Fidelity.}
The reconstructed modules are compared against the original ClinicalTrials.gov source via three complementary approaches.

\textbf{Human and LLM-based review.} Six trials were selected via stratified random sampling from the 23-trial corpus to ensure diverse intervention types and varying degrees of temporal complexity. Four TEL domain experts and a separate LLM instance (Gemini 3.1 Pro thinking, in a context-cleared session) independently assessed each reconstruction against the original source using four questions: semantic identity of temporal events (Q1), accuracy of temporal constraints (Q2), preservation of chronological requirements (Q3), and clarity of temporal representation (Q4). Each question was rated on a four-point scale (1 = completely identical/accurate/preserved/precise; 4 = not at all). Inter-rater reliability was measured using Fleiss' $\kappa$, and discrepancies among experts were resolved through consensus discussion.

\textbf{Algorithmic lexical and semantic scoring.} Algorithmic lexical and semantic scoring is applied independently across the six target protocol modules. For each module, all values are extracted from the JSON object with keys excluded, and non-string values such as booleans and integers are converted to their string equivalents before concatenation into a single text representation per module. We calculate two similarity metrics between the original source record and the TEL-derived reconstruction for each module. Lexical similarity is measured as TF-IDF weighted cosine similarity based on bag-of-words. Semantic similarity is computed by obtaining vector representations via the text-embedding-3-large model through the Azure OpenAI API. Both metrics are computed at the module level and macro-averaged across all trials to produce corpus-level fidelity scores.

To contextualize pairwise fidelity within the broader trial corpus, we generate a cross-trial similarity matrix for each module. In each matrix, rows correspond to original source records and columns to the TEL-derived reconstructions across all trial pairs. The diagonal entries capture paired translation fidelity which is the overall semantic similarity between each original trial and its own reconstruction. The off-diagonal entries serve as a natural baseline, which reflects the background similarity between an original record and the reconstruction of a different, unrelated trial. Because all trials are drawn from the same ``Alzheimer Disease and Related Dementias'' corpus and share the same protocolSection schema, this off-diagonal baseline accounts for domain-level similarity arising from shared clinical terminology, structural conventions, and ontology-mapped vocabulary. The degree to which diagonal scores exceed this baseline provides a discriminative measure of representational fidelity.

\section{Results}
\textbf{Experimental Setup.} 
Trial protocols were retrieved from ClinicalTrials.gov using a structured search restricted to Alzheimer Disease and Related Dementias (condition), Medication (intervention type), and older adults (age filter), with enrollment status limited to ''Active, Not Recruiting'' (n = 4) and ``Completed'' (n = 19) to ensure fully defined, stable protocol text across all six JSON modules. Medication protocols in this population consistently include well-defined temporal structures such as dosing schedules, follow-up windows, and cognitive outcome assessments that exercise the full range of TEL operators. The full search query is available at ClinicalTrials.gov\cite{clinicaltrials2026adrd}. All automated parsing, ontology mapping, and logical formulation steps were executed using the Gemini 3.1 Pro large language model.

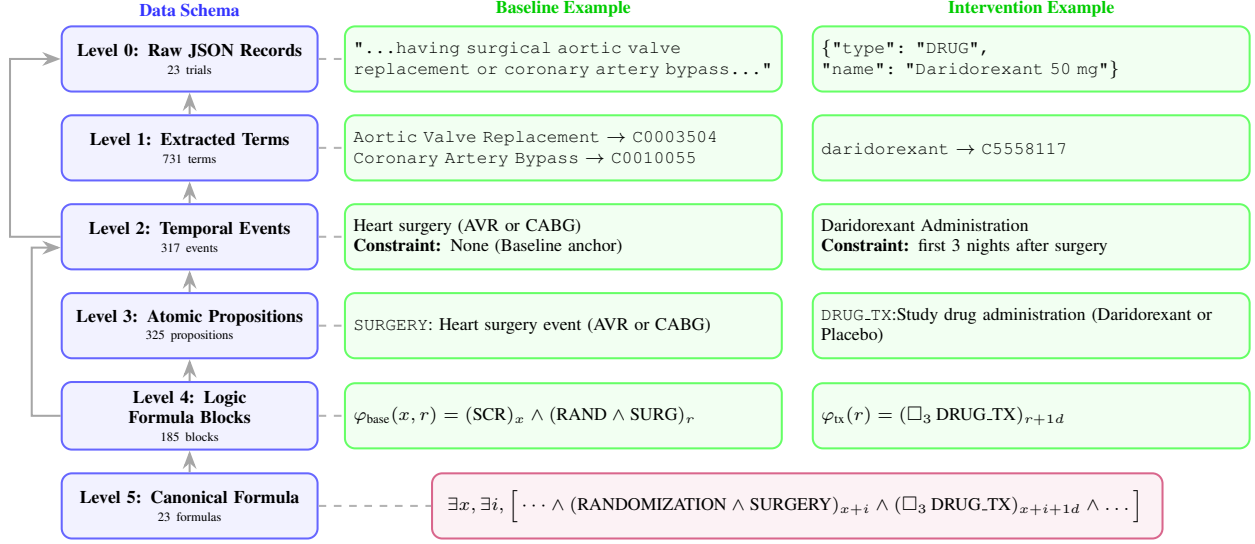
\begin{figure}[h!]
\centering
\resizebox{\linewidth}{!}{%
\begin{tikzpicture}[
    node distance=0.3cm and 0.4cm,
    schema/.style={rectangle, draw=blue!60, fill=blue!5, thick, rounded corners, text width=3.3cm, align=center, minimum height=0.9cm, font=\scriptsize\bfseries},
    example/.style={rectangle, draw=green!60, fill=green!5, thick, rounded corners, text width=5.8cm, align=left, minimum height=0.9cm, font=\scriptsize},
    level5/.style={rectangle, draw=purple!60, fill=purple!5, thick, rounded corners, align=center, minimum height=0.9cm, font=\scriptsize, inner sep=6pt},
    arrow/.style={-{Stealth[scale=1.1]}, thick, draw=gray!70}
]

\node[schema] (l0) {Level 0: Raw JSON Records\\[-1pt]{\mdseries\tiny 23 trials}};
\node[schema, below=of l0] (l1) {Level 1: Extracted Terms\\[-1pt]{\mdseries\tiny 731 terms}};
\node[schema, below=of l1] (l2) {Level 2: Temporal Events\\[-1pt]{\mdseries\tiny 317 events}};
\node[schema, below=of l2] (l3) {Level 3: Atomic Propositions\\[-1pt]{\mdseries\tiny 325 propositions}};
\node[schema, below=of l3] (l4) {Level 4: Logic Formula Blocks\\[-1pt]{\mdseries\tiny 185 blocks}};
\node[schema, below=of l4] (l5) {Level 5: Canonical Formula\\[-1pt]{\mdseries\tiny 23 formulas}};

\node[example, right=0.35cm of l0] (b0) {\texttt{"\dots having surgical aortic valve replacement or coronary artery bypass\dots"}};
\node[example, right=0.35cm of l1] (b1) {\texttt{Aortic Valve Replacement $\to$ C0003504}\\ \texttt{Coronary Artery Bypass $\to$ C0010055}};
\node[example, right=0.35cm of l2] (b2) {Heart surgery (AVR or CABG)\\ \textbf{Constraint:} None (Baseline anchor)};
\node[example, right=0.35cm of l3] (b3) { \texttt{SURGERY}: Heart surgery event (AVR or CABG)};
\node[example, right=0.35cm of l4] (b4) { $\varphi_{\text{base}}(x, r) = (\text{SCR})_{x} \land (\text{RAND} \land \text{SURG})_{r}$};

\node[example, right=of b0] (i0) {\texttt{\{"type": "DRUG",\\ "name": "Daridorexant 50 mg"\}}};
\node[example, right=of b1] (i1) {\texttt{daridorexant $\rightarrow$ C5558117}};
\node[example, right=of b2] (i2) {Daridorexant Administration\\ \textbf{Constraint:} first 3 nights after surgery};
\node[example, right=of b3] (i3) {\texttt{DRUG\_TX}:Study drug administration (Daridorexant or Placebo)};
\node[example, right=of b4] (i4) { $\varphi_{\text{tx}}(r) = (\Box_{3}\, \text{DRUG\_TX})_{r+1d}$};

\coordinate (e5mid) at ($(b4)!0.5!(i4)$);
\node[level5, anchor=center] (e5) at (e5mid |- l5) {
$\displaystyle \exists x, \exists i, \Big[ \dots \land (\text{RANDOMIZATION} \land \text{SURGERY})_{x+i} \land (\Box_{3} \, \text{DRUG\_TX})_{x+i+1d} \land \dots \Big]$
};

\draw[dashed, draw=gray!60, thick] (l0) -- (b0);
\draw[dashed, draw=gray!60, thick] (l1) -- (b1);
\draw[dashed, draw=gray!60, thick] (l2) -- (b2);
\draw[dashed, draw=gray!60, thick] (l3) -- (b3);
\draw[dashed, draw=gray!60, thick] (l4) -- (b4);
\draw[dashed, draw=gray!60, thick] (l5) -- (e5);

\draw[arrow] (l5) -- (l4);
\draw[arrow] (l4) -- (l3);
\draw[arrow] (l3) -- (l2);
\draw[arrow] (l2) -- (l1);
\draw[arrow] (l1) -- (l0);

\draw[arrow] (l4.west) -- ++(-0.4,0) |- ([yshift=-0.15cm]l2.west);
\draw[arrow] (l2.west) -- ++(-0.7,0) |- (l0.west);

\node[above=0.0cm of l0, font=\bfseries\scriptsize, text=blue!80] {Data Schema};
\node[above=0.0cm of b0, font=\bfseries\scriptsize, text=green!80!black] {Baseline Example};
\node[above=0.0cm of i0, font=\bfseries\scriptsize, text=green!80!black] {Intervention Example};

\end{tikzpicture}
}
\caption{Hierarchical data schema of the CT-TEL Library with aggregate counts (left) and two worked examples traced through all six levels (right). Arrows denote backward traceability.}
\label{fig:library_schema}
\end{figure}

\textbf{The CT-TEL Library.} Processing the 23 trials produced the CT-TEL Library, organized as a six-tier hierarchical data schema (Figure~\ref{fig:library_schema}) that preserves full provenance from raw text to formal logic. The tiers progress from raw JSON records (Level~0), through ontology-mapped terms (Level~1), temporally constrained events (Level~2), atomic propositions (Level~3), and modular logic blocks (Level~4), to the canonical TEL formula (Level~5), with explicit backward traceability at each tier. Across all 23 trials, the workflow produced 731 ontology-mapped terms, 317 temporal events, 325 atomic propositions, and 185 modular logic blocks, confirming that the workflow scales across diverse trial designs. Figure~\ref{fig:library_schema} traces two representative examples through all six levels using a baseline surgery event and a medication intervention.

\textbf{Example: From Canonical Formula to Protocol Narrative.}
To illustrate how the generated TEL formulas map back to interpretable clinical language, we present the canonical formula produced for trial NCT04179721, a cluster-randomized study evaluating a staff-training intervention (PES-4-BPSD) for behavioral symptoms of dementia. The workflow decomposed the protocol into four modular formula blocks. The \emph{eligibility block} encodes staff-level inclusion and exclusion criteria:
\[
\varphi_{\text{elig}} = (\textit{STAFF\_INC})_x \;\land\; (\neg\,\textit{DECLINE\_EXC})_x \;\land\; (\neg\,\textit{TEMP\_STAFF\_EXC})_x
\]
The \emph{baseline block} asserts that the target condition and baseline measures are recorded at enrollment~($x$):
\[
\varphi_{\text{base}} = (\textit{ADRD} \;\land\; \textit{BPSD} \;\land\; \textit{ADQ\_BL} \;\land\; \textit{SEWDR\_BL})_x
\]
The \emph{intervention block} uses the box operator $\Box_{3m}$ to express continuous 3-month training under either condition:
\[
\varphi_{\text{tx}} = (\Box_{3m}\;\textit{PES\_TX})_x \;\lor\; (\Box_{3m}\;\textit{CTRL\_TX})_x
\]
Finally, the \emph{outcome blocks} place follow-up assessments at 3-month and 15-month offsets using the diamond operator $\Diamond_{1m}$ for a measurement window:
\[
\varphi_{\text{out}} = (\Diamond_{1m}\,(\textit{ADQ\_3M} \land \textit{SEWDR\_3M}))_{x+3m} \;\land\; (\Diamond_{1m}\,(\textit{ADQ\_15M} \land \textit{SEWDR\_15M}))_{x+15m}
\]
These blocks are unified into a single canonical TEL formula:
\begin{equation}
\label{eq:nct04179721}
\Phi = \exists x \Big[\;
  \varphi_{\text{elig}} \;\land\; \varphi_{\text{base}} \;\land\; \varphi_{\text{tx}} \;\land\; \varphi_{\text{out}}
\;\Big]
\end{equation}

Reading this formula as natural language: \emph{``There exists a baseline time-point~$x$ at which permanently based nursing staff who have not declined participation are enrolled (eligibility). At~$x$, the presence of ADRD and BPSD is confirmed and baseline ADQ and SEWDR scores are recorded. Starting at~$x$, staff undergo either the PES-4-BPSD training or the attention-control training continuously for 3~months. Within a 1-month window after the 3-month mark, ADQ and SEWDR are reassessed; and again within a 1-month window after the 15-month mark.''}
This back-translation demonstrates that every temporal operator in the formula ($\Box$ for continuous phases, $\Diamond$ for measurement windows, subscript offsets for scheduled time-points) has a direct, unambiguous clinical interpretation, confirming the semantic fidelity of the automated formalization.

\subsection{Evaluation}
\textbf{Human and LLM-Based Review.} The six sampled trials were NCT00104442, NCT04179721, NCT05741060, NCT04308512, NCT06335953, and NCT06630390. Fleiss' $\kappa$ among the four experts was $\text{0.548}$, indicating moderate agreement. Across $\text{24}$ evaluated items (6 trials $\times$ 4 questions), the consensus human mean was $\text{1.89}$ and the LLM mean was $\text{1.85}$. Q1 (human $\text{2.08}$, LLM $\text{2.0}$) and Q2 (human $\text{1.92}$, LLM $\text{2.0}$) show temporal events and constraints are systematically captured, with occasional partial mismatch or omission. Q3 (human $\text{1.92}$, LLM $\text{2.17}$) confirmed chronological sequencing is maintained, though the LLM rater registered slightly more divergence. Q4 returned the strongest scores (human $\text{1.63}$, LLM $\text{1.33}$), indicating reconstructed protocols are temporally precise and unambiguous.

\begin{figure}[h]
\centering
\includegraphics[width=1\textwidth]{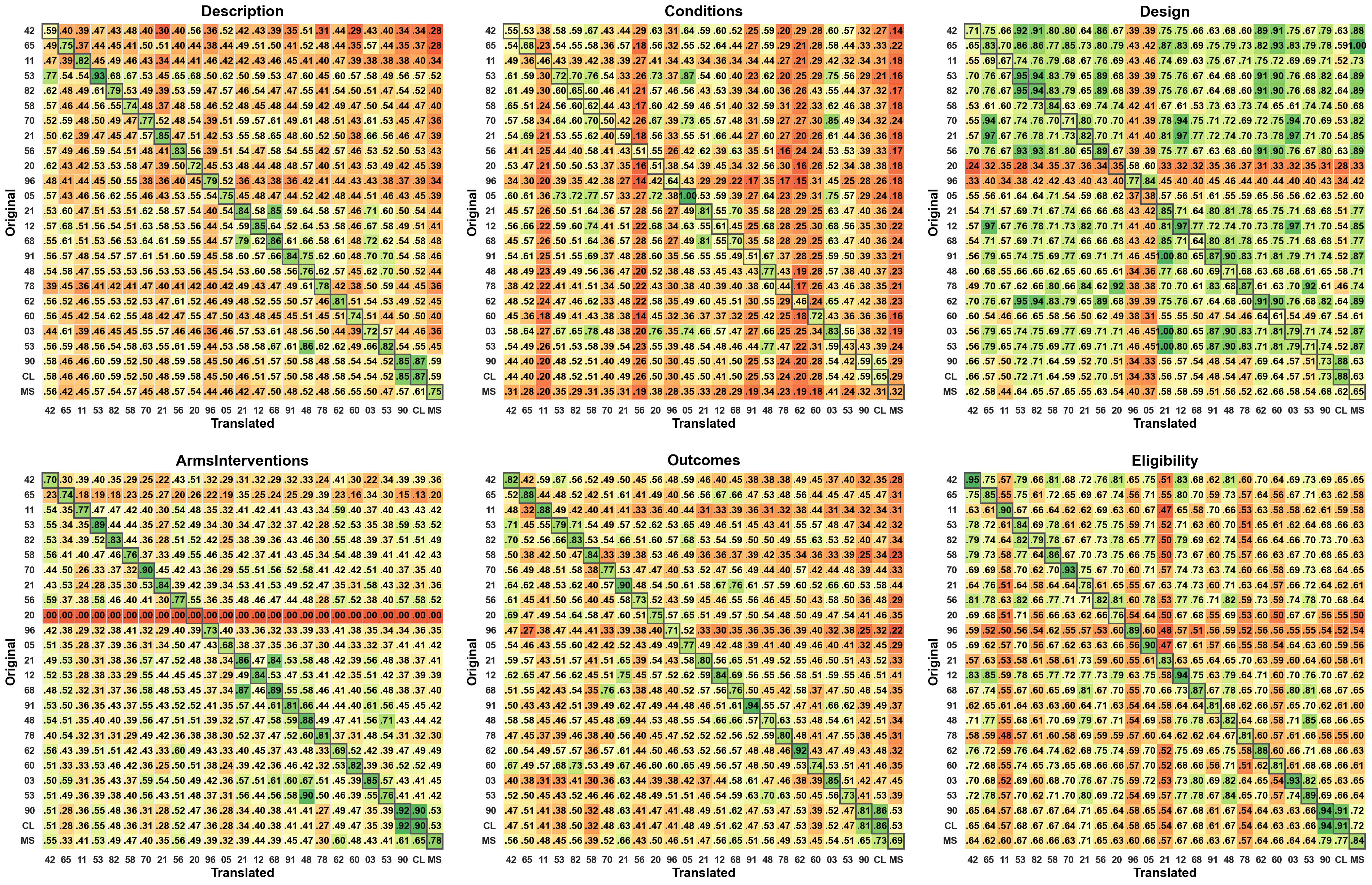}
\caption{Cross-trial semantic similarity per module. Rows correspond to original records and columns to translations. Axis labels show the last two digits of each NCT identifier, e.g., 90 corresponds to NCT06630390; CL denotes NCT06630390\textsubscript{Claude} and MS denotes NCT06630390\textsubscript{masked}. Higher diagonal scores relative to the off-diagonal background indicate preservation of trial-specific semantic content.}
\label{heatmap}
\end{figure}

\textbf{Algorithmic Lexical and Semantic Scoring.}
Table~\ref{tab:similarity} reports lexical and semantic similarity scores across all trial pairs and six protocol modules. Semantic scores are consistently higher than lexical scores: averaged across all trials, semantic scores range from 0.622 (conditions) to 0.860 (eligibility), while lexical scores range from 0.363 (conditions) to 0.460 (eligibility). The conditions module scores lowest under both metrics, likely because its concise, terminology-dense content means minor phrasing variation penalizes both measures. Embedding-based metrics cannot detect logic errors such as collapsed duration states, hallucinated fields, or incorrect temporal anchors; logic-level validation via model-checking against the same Temporal Event Structure would catch what surface similarity misses.
The cross-trial heatmaps in Figure~\ref{heatmap} confirm that diagonal entries (each trial against its own reconstruction) are visibly elevated above off-diagonal entries, which capture background similarity across all ADRD trial records. This separation shows the reconstructed protocols are faithful to their specific source trials, not just reproducing shared clinical trial vocabulary.
Table~\ref{tab:trial_translation_similarity} shows five module-level examples. NCT03901105 achieves perfect scores (1.000/1.000) on a single-label conditions module. NCT04308512 scores 0.847/0.966 on design, with the reconstruction adding a \texttt{maskingInfo} field absent from the original, a minor hallucination that preserves high semantic fidelity. NCT04481568 scores 0.386/0.889 on armsInterventions: the reconstruction captures the correct clinical intent but condenses a detailed description into a shorter paraphrase, producing low lexical overlap alongside strong semantic retention.

\begin{table}[t]
\centering
\caption{Lexical and semantic similarity scores between the original and translated
by trial and module, reported as lexical\,/\,semantic. The average is computed
over the 23 canonical trials only; the two ablation variants of NCT06630390
are excluded from the macro-average.}
\label{tab:similarity}
\vspace{4pt}
\resizebox{\textwidth}{!}{%
\renewcommand{\arraystretch}{0.95}
\setlength{\tabcolsep}{5pt}
\small
\begin{tabular}{lcccccc}
\toprule
\textbf{Trial}
  & \textbf{Description}
  & \textbf{Conditions}
  & \textbf{Design}
  & \textbf{ArmsInterventions}
  & \textbf{Outcomes}
  & \textbf{Eligibility} \\
\midrule
NCT00104442 & 0.386\,/\,0.588 & 0.401\,/\,0.553 & 0.354\,/\,0.709 & 0.280\,/\,0.703 & 0.517\,/\,0.818 & 0.401\,/\,0.951 \\
NCT00178165 & 0.445\,/\,0.747 & 0.448\,/\,0.679 & 0.580\,/\,0.834 & 0.498\,/\,0.737 & 0.479\,/\,0.881 & 0.662\,/\,0.848 \\
NCT00870311 & 0.423\,/\,0.815 & 0.206\,/\,0.461 & 0.335\,/\,0.665 & 0.596\,/\,0.767 & 0.214\,/\,0.881 & 0.204\,/\,0.900 \\
NCT01428453 & 0.481\,/\,0.929 & 0.803\,/\,0.721 & 0.847\,/\,0.954 & 0.354\,/\,0.888 & 0.495\,/\,0.786 & 0.701\,/\,0.843 \\
NCT02646982 & 0.270\,/\,0.785 & 0.591\,/\,0.651 & 0.847\,/\,0.941 & 0.307\,/\,0.836 & 0.497\,/\,0.831 & 0.499\,/\,0.789 \\
NCT02707458 & 0.393\,/\,0.742 & 0.367\,/\,0.615 & 0.503\,/\,0.843 & 0.271\,/\,0.764 & 0.586\,/\,0.836 & 0.601\,/\,0.858 \\
NCT02833870 & 0.477\,/\,0.771 & 0.136\,/\,0.498 & 0.252\,/\,0.706 & 0.692\,/\,0.896 & 0.453\,/\,0.766 & 0.570\,/\,0.935 \\
NCT03046121 & 0.550\,/\,0.850 & 0.041\,/\,0.592 & 0.439\,/\,0.818 & 0.470\,/\,0.845 & 0.724\,/\,0.899 & 0.401\,/\,0.782 \\
NCT03325556 & 0.362\,/\,0.827 & 0.236\,/\,0.510 & 0.439\,/\,0.886 & 0.257\,/\,0.770 & 0.270\,/\,0.736 & 0.392\,/\,0.817 \\
NCT03349320 & 0.352\,/\,0.715 & 0.392\,/\,0.510 & 0.000\,/\,0.349 & 0.000\,/\,0.000 & 0.256\,/\,0.755 & 0.429\,/\,0.760 \\
NCT03431896 & 0.359\,/\,0.787 & 0.166\,/\,0.644 & 0.510\,/\,0.769 & 0.286\,/\,0.727 & 0.354\,/\,0.713 & 0.309\,/\,0.888 \\
NCT03901105 & 0.260\,/\,0.754 & 1.000\,/\,1.000 & 0.000\,/\,0.384 & 0.374\,/\,0.679 & 0.265\,/\,0.767 & 0.814\,/\,0.896 \\
NCT04179721 & 0.644\,/\,0.843 & 0.932\,/\,0.814 & 0.336\,/\,0.853 & 0.441\,/\,0.859 & 0.492\,/\,0.800 & 0.467\,/\,0.827 \\
NCT04308512 & 0.417\,/\,0.848 & 0.121\,/\,0.615 & 0.847\,/\,0.966 & 0.433\,/\,0.837 & 0.301\,/\,0.844 & 0.338\,/\,0.938 \\
NCT04481568 & 0.631\,/\,0.864 & 0.394\,/\,0.696 & 0.089\,/\,0.643 & 0.386\,/\,0.889 & 0.618\,/\,0.759 & 0.404\,/\,0.868 \\
NCT04851691 & 0.514\,/\,0.836 & 0.178\,/\,0.515 & 0.510\,/\,0.870 & 0.369\,/\,0.806 & 0.800\,/\,0.939 & 0.424\,/\,0.807 \\
NCT04938648 & 0.535\,/\,0.757 & 0.407\,/\,0.768 & 0.155\,/\,0.713 & 0.609\,/\,0.882 & 0.315\,/\,0.697 & 0.217\,/\,0.820 \\
NCT05535478 & 0.466\,/\,0.785 & 0.096\,/\,0.443 & 0.674\,/\,0.870 & 0.495\,/\,0.805 & 0.684\,/\,0.797 & 0.266\,/\,0.808 \\
NCT05644262 & 0.560\,/\,0.811 & 0.306\,/\,0.458 & 0.519\,/\,0.915 & 0.443\,/\,0.690 & 0.482\,/\,0.925 & 0.494\,/\,0.882 \\
NCT05741060 & 0.500\,/\,0.743 & 0.494\,/\,0.720 & 0.275\,/\,0.613 & 0.346\,/\,0.821 & 0.239\,/\,0.738 & 0.407\,/\,0.809 \\
NCT06079203 & 0.327\,/\,0.715 & 0.533\,/\,0.825 & 0.580\,/\,0.795 & 0.567\,/\,0.852 & 0.467\,/\,0.847 & 0.459\,/\,0.927 \\
NCT06335953 & 0.527\,/\,0.817 & 0.000\,/\,0.426 & 0.322\,/\,0.711 & 0.287\,/\,0.756 & 0.298\,/\,0.736 & 0.469\,/\,0.887 \\
NCT06630390 & 0.517\,/\,0.850 & 0.099\,/\,0.590 & 0.295\,/\,0.728 & 0.605\,/\,0.922 & 0.454\,/\,0.808 & 0.647\,/\,0.941 \\
\midrule
\textbf{Average (23 trials)}
            & \textbf{0.452\,/\,0.790} & \textbf{0.363\,/\,0.622} & \textbf{0.422\,/\,0.762} & \textbf{0.407\,/\,0.771} & \textbf{0.446\,/\,0.807} & \textbf{0.460\,/\,0.860} \\
\midrule
\multicolumn{7}{l}{\textit{LLM sensitivity ablations (excluded from average)}} \\[2pt]
NCT06630390\textsubscript{Claude}  & 0.535\,/\,0.866 & 0.239\,/\,0.648 & 0.252\,/\,0.878 & 0.450\,/\,0.898 & 0.559\,/\,0.857 & 0.553\,/\,0.906 \\
NCT06630390\textsubscript{masked}  & 0.493\,/\,0.754 & 0.000\,/\,0.320 & 0.099\,/\,0.644 & 0.645\,/\,0.777 & 0.296\,/\,0.692 & 0.555\,/\,0.845 \\
\bottomrule
\end{tabular}%
}
\end{table}

\begin{table}[h]
\centering
\caption{Five examples comparing original and translated clinical trial module contents.}
\label{tab:trial_translation_similarity}
\resizebox{\textwidth}{!}{%
\small
\renewcommand{\arraystretch}{1.0}
\setlength{\tabcolsep}{4pt}
\begin{tabular}{@{} >{\RaggedRight}p{1.7cm}
    >{\RaggedRight\ttfamily}p{9cm}
    >{\RaggedRight\ttfamily}p{9.8cm}
    >{\centering\arraybackslash}p{1.3cm}}
\toprule
\normalfont\textbf{Trial\ \ \  (Module)} &
\normalfont\textbf{Original Content} &
\normalfont\textbf{Translated Content} &
\normalfont\textbf{Sim.\ (Lex\,/\,Sem)} \\
\midrule

NCT03901105 (Conditions)
  & \{"conditions": ["Alzheimer Disease"]\}
  & \{"conditions": ["Alzheimer Disease"]\}
  & 1.0000\,/ \,1.0000 \\
\addlinespace[3pt]

NCT04308512 (Design)
  & \{"studyType": "INTERVENTIONAL", "phases": ["NA"], "designInfo": \{"allocation": "RANDOMIZED", "interventionModel": "PARALLEL", "primaryPurpose": "SUPPORTIVE\_CARE"\}\}
  & \{"studyType": "INTERVENTIONAL", "phases": ["NA"], "designInfo": \{"allocation": "RANDOMIZED", "interventionModel": "PARALLEL", "primaryPurpose": "SUPPORTIVE\_CARE", "maskingInfo": \{"masking": "NONE"\}\}\}
  & 0.8466\,/ \,0.9661 \\
\addlinespace[3pt]

NCT02833870 (Eligibility)
  & \{"eligibilityCriteria": "Inclusion: Older francophones in institutions, cortical degenerative disease, age 60--99, French mother tongue, social security scheme, participation agreement. Exclusion: No cortical degenerative disease, uncorrected profound hearing loss, not affiliated with social security scheme."\}
  & \{"eligibilityCriteria": "Inclusion: Cortical degenerative disease, age 60--99, older francophones in institutions, participation agreement. Exclusion: Absence of cortical degenerative disease, uncorrected profound hearing loss, not affiliated with social security scheme.", "healthyVolunteers": false, "sex": "ALL", "minimumAge": "60 Years", "maximumAge": "99 Years"\}
  & 0.5698\,/ \,0.9345 \\
\addlinespace[3pt]

NCT04481568 (ArmsInterventions)
  & \{"armGroups": [\{"label": "The PES-4-BPSD Model", "type": "EXPERIMENTAL", "description": "Intervention arm of PES and nurse assistants: cohorting of patients with cognitive impairment AND BPSD, acutely admitted to medicine or telemetry service, on a 10-bed medical unit."\}]\}
  & \{"armGroups": [\{"label": "PES-4-BPSD Intervention", "type": "Experimental", "description": "Patients receive PES-4-BPSD model continuously with daily behavioral and delirium monitoring."\}, \{"label": "Attention Control", "type": "Active Comparator", "description": "Patients receive attention control continuously with daily behavioral and delirium monitoring."\}]\}
  & 0.3861\,/ \,0.8893 \\
\addlinespace[3pt]

NCT04851691 (Outcomes)
  & \{"primaryOutcomes": [\{"measure": "Cumulative Total of New Antipsychotic Pill-days Prescribed", "description": "Cumulative total of new antipsychotic prescription days per eligible patient in the 12 months after intervention rollout vs.\ prior 12 months", "timeFrame": "12 month time point"\}]\}
  & \{"primaryOutcomes": [\{"measure": "Cumulative Total of New Antipsychotic Pill-days Prescribed", "description": "The cumulative number of new antipsychotic pill-days prescribed, measured during the 12 months following the intervention.", "timeFrame": "Up to 12 months after intervention"\}]\}
  & 0.8002\,/ \,0.9386 \\

\bottomrule
\end{tabular}
}
\end{table}

\section{Discussion}

\subsection{LLM Sensitivity and Ontology Masking}
We compared three variants of trial NCT06630390: the default Gemini 3.1 Pro thinking translation, a Claude Sonnet 4.6 thinking-augmented translation, and an ontology-masked variant in which all mapped ontology terms were replaced with opaque labels (e.g., \texttt{Disease-02}, \texttt{Drug-24}).

\textbf{Cross-LLM comparison.}
As shown in Table~\ref{tab:similarity} row \textit{NCT06630390} and \textit{NCT06630390-claude}, the Claude-based translation achieved comparable fidelity, outperforming Gemini on semantic similarity in four of six modules while falling slightly behind on armsInterventions and eligibility. This suggests the CT-TEL workflow generalizes across capable reasoning models, though identifying the optimal LLM requires further investigation. Notably, OpenAI ChatGPT 5.4 failed to produce valid output with the same prompt, indicating that not all frontier models possess the instruction-following precision required for formal logic synthesis.

\textbf{Concept-term masking experiment.}
Row \textit{NCT06630390} and \textit{NCT06630390-masked} in Table~\ref{tab:similarity} show that the masked variant performed worse across all six modules. By stripping meaningful ontology labels, the reverse-translation LLM loses clinical context needed to reconstruct text resembling the original. However, this does not necessarily mean the underlying \textit{logical structure} is less accurate because the masked formulas may still encode correct temporal operators, modal scoping, and Boolean dependencies. The fidelity drop is partly an artifact of evaluation metrics that rely on surface-level similarity. 
Future work should incorporate logic-level evaluation, such as formula equivalence checking via canonical normalization, or model-checking-based validation where both original and reconstructed formulas are executed against the same TES and their satisfaction sets compared, directly measuring whether formalized logic preserves operational semantics independent of surface form.

\subsection{Downstream Applications}
Translating narrative protocols into TEL at scale unlocks robust downstream applications in CTS. Using the model-checking algorithms established for QEL \cite{qel}, these formulas can be executed directly against TES derived from real-world electrophysiology data or EHRs. This facilitates:
1. Eligibility Cohort Building Simulation: Formal logic allows databases to be queried to model virtual cohorts, adjusting temporal constraints to optimize recruitment strategies \cite{time_tel_cts};
2. Self-Controlled Case Series (SCCS): By representing interventions and outcomes in TEL, synthetic control arms can be dynamically constructed by establishing individualized timelines for real-world patients.

\subsection{Error Analysis of Temporal Extractions}

Two primary failure modes were identified during the formalization of temporal structures.

\textbf{Temporal distortion of nested constraints.} The pipeline occasionally struggles with highly complex, multi-stage longitudinal constraints, simplifying or subtly altering conditions during formalization. 
For example, a complex dose-escalation schedule requiring monitoring over overlapping sliding windows may be incorrectly collapsed into a single, contiguous duration state ($\Box_t$), because the LLM conflates the fine-grained hierarchical boundaries of overlapping time structures within its context window.

\textbf{Implicit assumption injection.} When protocol narratives leave logical gaps, the LLM occasionally fills them with standard clinical conventions not explicitly stated in the source text. A representative case is the design module 
for NCT04308512 (Table~\ref{tab:trial_translation_similarity}), where the reconstructed output introduced a \texttt{maskingInfo} field entirely absent from the original protocol. 
This occurs because the LLM attempts to satisfy the complete structure required by the JSON schema, rather than outputting a null or empty state for missing fields.

\subsection{Limitations and Future Work}
While this systematic approach successfully scales the formal specification of protocols, the current implementation relies heavily on the underlying reasoning capacity of the LLM. Future iterations will explore fine-tuning open-source LLMs strictly on TEL-to-protocol corpora to eliminate dependence on proprietary APIs. We also aim to directly integrate automated model-checking engines to instantly query OMOP-formatted EHR databases using the synthesized TEL formulas, bridging the gap between symbolic timing constraints and standardized clinical vocabularies.

\section{Conclusion}

The reliance on free-text narrative for clinical trial protocols has long impeded automated computation over trial design. The CT-TEL workflow addresses this by leveraging LLMs to translate narrative protocols into Temporal Ensemble Logic at scale. Round-trip semantic evaluation across 23 real-world trials confirmed that complex clinical timelines can be faithfully rendered into computable logical structures, with semantic similarity scores ranging from 0.622 (conditions module) to 0.860 (eligibility module).
This work contributes a foundational component to the ``Symbolic Biomedicine'' paradigm~\cite{qel, zhang2024temporal}, which advocates harmonizing data-driven neural AI with the verifiability of rule-based logical reasoning. 
CT-TEL operationalizes this vision: machine-readable, temporally aware protocol models can serve as direct inputs to model-checking engines for eligibility cohort simulation and self-controlled case series analysis against real-world EHR data.

\section*{Acknowledgments}
This work was supported in part by the National Science Foundation Award IIS2500624 and the National Institutes of Health grants R01AG084236 and U24AG098157, of the United States. The views of the paper are those of the authors and do not reflect those of the funding agencies.

\makeatletter
\renewcommand{\@biblabel}[1]{\hfill #1.}
\makeatother

\centering
\bibliographystyle{vancouver}
\setlength\itemsep{0pt}

\end{document}